\documentclass[twocolumn]{aastex63}
\usepackage[utf8]{inputenc}
\usepackage{graphicx}

\newcommand{\lea}{{\>\rlap{\raise2pt\hbox{$<$}}\lower3pt\hbox{$\sim$} \>}}
\newcommand{\gea}{{\>\rlap{\raise2pt\hbox{$>$}}\lower3pt\hbox{$\sim$} \>}}

\begin{document}

\title{The Ancient Globular Clusters of NGC~1291} 

\correspondingauthor{Kyle Hixenbaugh}
\author{Kyle Hixenbaugh} 
\author{Rupali Chandar}
\affiliation{Ritter Astrophysical Research Center, The University of Toledo, Toledo, OH 43606, USA}
\author{Angus Mok}
\affiliation{OCAD University, Toronto, Ontario, M5T 1W1, Canada}

\begin{abstract}
We present a new catalog 
of 81 ancient globular clusters (GCs) in the early-type spiral (SB0/a) galaxy NGC~1291. Candidates have been selected from $B$,$V$, and $I$ band images taken with the Hubble Space Telescope, which also reveal 17 younger ($\tau \lea \mbox{few}\times100$~Myr) clusters. The luminosity function shows a peaked shape similar to that found for GC systems in other spiral and elliptical galaxies. The ancient clusters have a bimodal color distribution, with approximately 65\% (35\%) of the population having blue (red) colors. The red, presumably metal-rich GCs are more centrally concentrated, as expected for a bulge population; while the blue, presumably metal-poor GCs, are more broadly distributed, consistent with expectations of a halo population. The specific frequency of GCs in NGC~1291 is higher than found previously in most spiral galaxies.  However, if we consider just the blue subpopulation, we find $S_{\rm N, blue}=0.50\pm0.06$, quite similar to that found for other spirals. This result supports the hypothesis of a universal population of halo GCs in spirals.  The fraction of red GCs in NGC~1291, when compared with those found in other galaxies, suggests that these correlate with host galaxy type rather than with host galaxy luminosity.
\end{abstract}

\section{Introduction} \label{sec:intro}
Globular clusters (GCs) are some of the oldest known objects in the universe, and as such give clues to the early stages of galaxy formation.  These ancient ($\tau \gea 10$~Gyr), massive ($M \approx 10^4 - 10^6~M_{\odot}$) stellar systems \citep{Brodie} are found in the halos of nearly all galaxies, at least those with stellar masses greater than $\sim10^9~M_{\odot}$ \citep{Harris17}. In earlier-type galaxies, some globular clusters are believed to be associated with bulges \citep{Zinn}, and in some lower mass spiral and irregular galaxies they have also been shown to have disk kinematics \citep{Olsen,Schommer}.  

The Hubble Space Telescope (HST) has  
enabled a number of major breakthroughs in our understanding of globular cluster systems, primarily in elliptical galaxies, which have fairly simple morphologies and only contain ancient clusters \citep{Peng,Peng08,Jordan}. Spiral galaxies, by contrast, have more complicated structures due to on-going star-formation in their disks, and can have rich populations of young clusters, \citep{Zwart,Christian,Chandar99,Chandar01}, making it more challenging (but not impossible) to identify ancient globular clusters \citep{Chandar}. Outside of the Milky Way, M31 \citep{Huxor11, Huxor14} and M81 \citep{Nantais10,Nantais11,SC10} are the spiral galaxies with the best-studied globular cluster systems.  Other spirals which have more recent catalogs of globular clusters from the Hubble Space Telescope include: M101, M51, NGC~4258, and NGC~6278 \citep{Lomeli22}. 

Lenticular and early-type spirals provide an important transition between non-star forming ellipticals and later-type galaxies with significant current star formation.  Lenticular and early type galaxies have bimodal GC populations: one blue and metal-poor, and the other red and metal-rich, whose colors correlate with the host galaxy luminosity and color \citep{Peng}. This has been studied in galaxies such as the lenticular galaxy NGC~1172 \citep{Ennis1172} and the early-type NGC~6876 \citep{Ennis6876}. These GC populations have distinct spatial distributions, with redder, metal rich GC's more centrally concentrated than their metal-poor counterparts. The distinct subpopulations, along with with high specific frequencies (the number of GCs normalized to a galaxy with $M_V=-15$), are believed to result from earlier merger events \citep{Ennis1172,Ennis6876}.

In this work, we present the first study of the globular cluster system of NGC~1291, an early-type barred spiral (SB0/a) galaxy that is forming stars slowly in its disk \citep{Thilker}, and has a prominent bulge \citep{Vaucouleurs}. The galaxy presents nearly face-on \citep{Bosma}. 
In this work, we adopt a distance of $8.9$~Mpc ($m-M=29.75)$ \citep{Vaucouleurs}, based on distance modulus estimations from bright stars, which is quite similar to the estimate of $9.08 \pm 0.29$~Mpc made by \citet{McQuinn} based on the tip of the red giant branch method.

We use archival, optical images of four pointings within NGC~1291 taken with $HST$ to identify and study it's globular cluster system. It should be noted the globular clusters identified in this work are candidates and have no velocity confirmation. For simplicity, we refer to these candidates as globular clusters throughout this paper.
The rest of this paper is organized as follows.  In Section~2 we present how we select our globular cluster sample, including photometry and an estimate of completeness.  Section~3 presents the color-magnitude and color-color distributions of the globular clusters, along with estimates of their masses. Section~4 focuses on the luminosity/mass functions, color/metallicity distributions, and spatial distributions of the ancient clusters.  We present and discuss the specific frequency of the GCs in Section~5, and summarize our main results in Section~6.

\section{Observations, Cluster Selection, and Photometry} \label{sec:data}
\subsection{Observations and Photometry}

\begin{figure}[ht!]
\includegraphics[width=\linewidth]{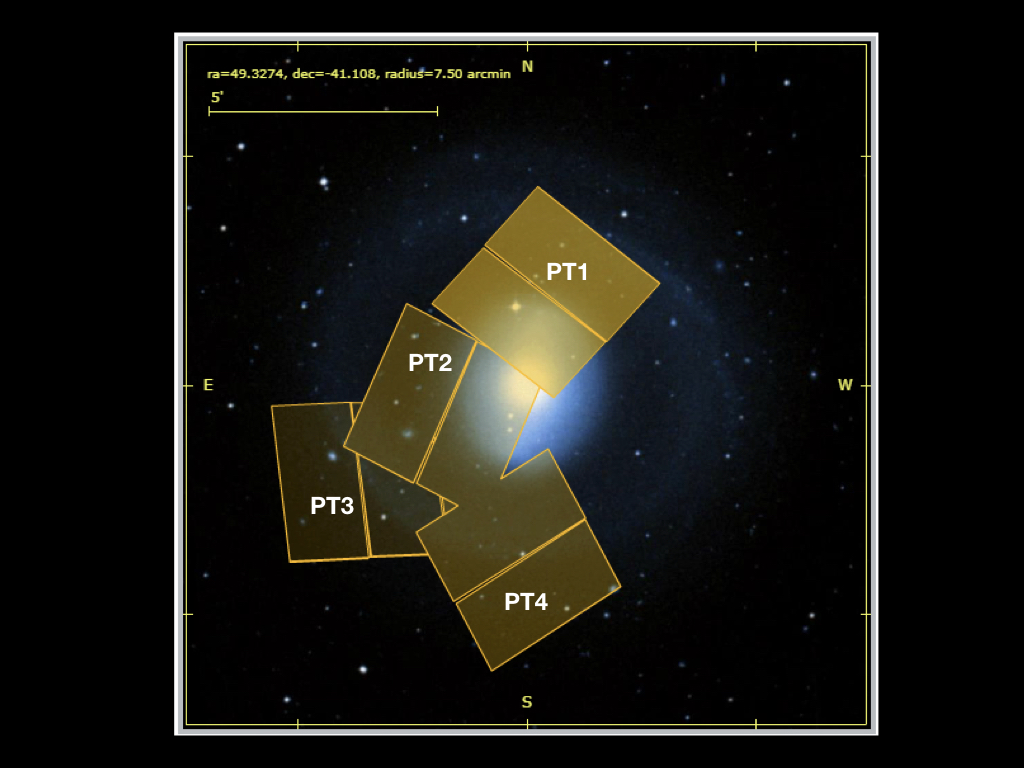}
\caption{Ground-based optical image of NGC~1291 showing the locations of the four $HST$ ACS/WFC pointings (labeled) used in this work. A $5\arcmin$ scale bar is shown for reference.
\label{fig:pts}}
\end{figure}

We use 4 pointings in NGC~1291 taken by the HST ACS/WFC camera in 2005 April and 2013 September.  All pointings have images in three filters: F435W ($B$ band), F555W or F606W ($V$ band), and F814 ($I$ band). Each field covers an area of approximately 40,000~$\arcsec^2$ or $74~\mbox{kpc}^2$. At the adopted distance of 8.9~Mpc to NGC~1291, the fields used cover a total area of around $296~\mbox{kpc}^2$,  and the pixel scale is 43~pc~arcsec$^{-1}$. Figure~\ref{fig:pts} shows the locations of the 4 pointings, and Table~\ref{tab:img} compiles basic information for each one, including position.

Reduced images were downloaded from the Hubble Legacy Archive (HLA).\footnote{https://hla.stsci.edu/hlaview.html} These images have been reduced through the HLA pipeline, which includes bias subtraction, dark subtraction, flat-fielding, cosmic ray rejection, and drizzling.

We detect sources using the DAOfind routine in the Astropy Python package. The detection algorithm is run on each $V$ band image, set to approximately a $3\sigma$ threshold. A visual inspection confirmed that essentially all obvious, point-like sources are detected. The list of detected objects includes star clusters, individual stars, and background galaxies. Masks are applied to bright, saturated foreground stars to prevent their detection. Aperture photometry was performed in each available filter using the photutils python package in Astropy. The brightness of these sources is measured using a circular aperture with radius of 3 pixels, with the background level estimated in an annulus with radii between 5 and 8 pixels.  We apply the following zeropoints (obtained from the STScI webpage) to our aperture photometry to put it on the VEGAMAG system: $F435W=25.788$, $F555W=25.732$, $F606W=26.406$, and $F814W=25.528$. Table~\ref{tab:cat} compiles the magnitudes, positions, and size information (described in Section~\ref{sec:sizes}) for five clusters in our catalog. The full table has released with this publication. The VEGAMAG photometric system is similar to the Johnson-Cousins system for the ACS/WFC filters used in this work, with only small differences in the zeropoints on the order of $\sim0.03-0.05$~mag \citep{Sirianni05}.  

In order to determine aperture corrections, we measure the curve of growth out to 20~pixels for ten isolated clusters in the $V$ band images. We find an average aperture correction (between 3 and 20 pixels) of $-0.47$~mag, and include an additional $-0.1$~mag to correct from 20 pixels to infinity, giving a total $V$-band aperture correction of $-0.57$~mag. This average correction was applied to each cluster (selection described below). In addition, PT3 in Figure~\ref{fig:pts} is observed in the F606W rather than the F555W filter. To account for this, we find 15 bright and isolated sources in the region of overlap between PT3 and PT2. We calculate the magnitude difference for the sources in common between the two filters, and find an average difference of $\sim 0.258$~mag. We add this to offset to the F606W photometry to convert to the F555W band.

In the rest of this work, we use magnitudes that have been corrected for the following foreground extinction values: $m_{\rm F435W}=0.047$~mag, $m_{\rm F555W}=0.036$~mag, and $m_{\rm F814W}=0.02$~mag, which gives small shifts in the colors of $m_{\rm F435W} - m_{\rm F555W}=0.011$~mag and $m_{\rm F555W} - m_{\rm F814W}=0.016$~mag \cite{Reddening}.\footnote{Extinction values can be found with NED (https://ned.ipac.caltech.edu), along with the paper producing the values, \cite{Reddening}.}

\subsection{Cluster Selection}
\begin{figure}[ht!]
\includegraphics[width=\linewidth]{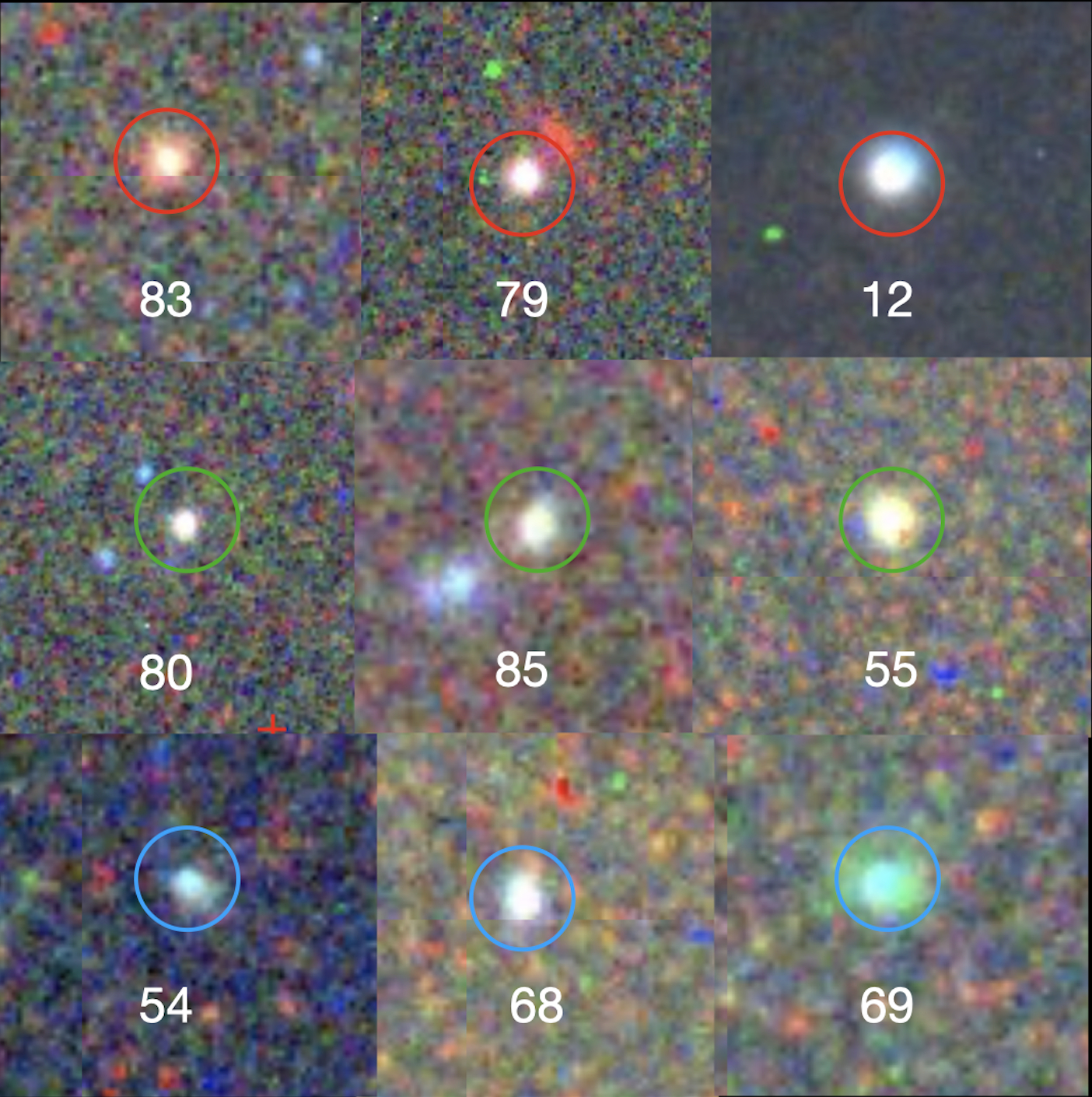}
\caption{Three color images of typical candidate clusters, showing the range of colors from reddest (top row), to intermediate (middle row), to bluest (bottom row) of the detected clusters. Each panel covers an approximate area of $~18"^{2}$, or $78~\mbox{pc}^{2}$ , and gives the corresponding ID number from our catalog.
\label{fig:stamps}}
\end{figure}

The initial source list contains many objects, including, clusters and bright, individual stars in NGC~1291, background galaxies, and foreground stars. The goal of this work is to select and study the ancient globular clusters in NGC~1291. To remove contaminants and select candidate ancient clusters, we use automated selection criteria followed by a final visual inspection. Given the relative proximity of NGC~1291, essentially all globular clusters are broader than the PSF  (see Section \ref{sec:sizes} for size measurements. We show some example clusters from our dataset in  Figure~\ref{fig:stamps}.

\begin{figure}[ht!]
\includegraphics[width=\linewidth]{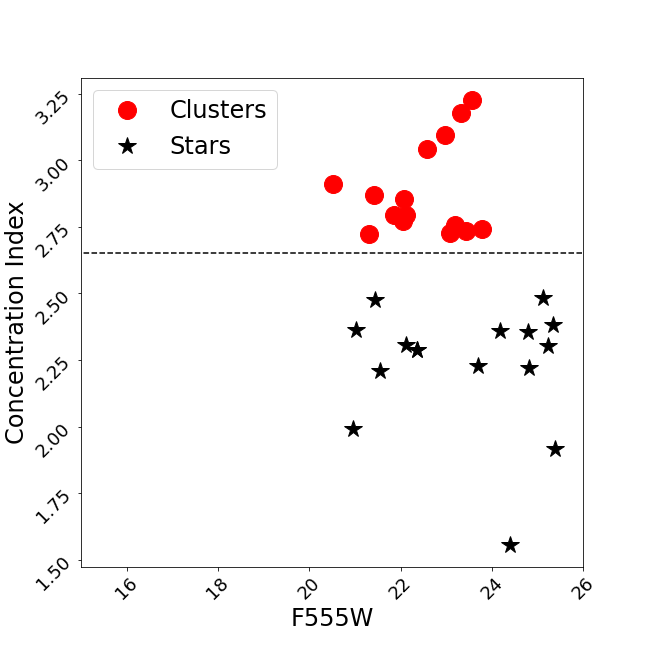}
\caption{Plot of concentration index (CI) vs apparent magnitude (both measured in F555W) for the training set of stars and clusters. There is a clear separation between the two objects. The black lines represent the CI$=2.65$ used to automatically select cluster candidates. 
\label{fig:ci}}
\end{figure}

We build training sets of $\sim15$ hand-selected stars, clusters, and galaxies to help guide the automated cuts. 
Following \citet{Whitmore21}, we identify our training set using a variety of tools, including color images and plotting radial profiles using the imexam python package, where stars all have similar sharp radial profiles but clusters are clearly more extended. 
Our primary quantitative tool is the 
Concentration Index (CI), which is the difference in the V-magnitude measured in apertures of 0.5 and 3 pixel radii. Clusters have a larger measured CI than stars. In Figure~\ref{fig:ci}, we plot the CI vs. V magnitude for our training set of stars (star symbols) and clusters (red circles). The clusters clearly have larger values of CI at similar magnitudes. Based on the measurements of the sources in our training set, we assume that sources with CI$>2.65$ are candidate clusters.

We eliminate all sources with $CI<2.65$, then visually inspect all remaining sources.  Background galaxies and close stellar pairs or triples, which also have larger values of CI, were eliminated by visual inspection. This procedure resulted in 98 candidate clusters; we show representative color images for these sources in Figure~\ref{fig:stamps}. Three clusters representing the range of cluster colors (which we describe in the next section), were selected to show the range of cluster appearances.  While we believe that our catalog of GCs is fairly contamination free, almost all cluster catalogs selected from imaging alone have at least some low-level of contamination from spurious sources, and future radial velocity measurements could be used to eliminate any contaminants.

\subsection{Completeness}
\label{sec:completeness}

\begin{figure}[ht!]
\includegraphics[width=\linewidth]{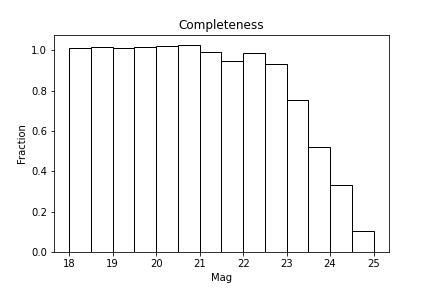}
\caption{Completeness fraction for artificial sources recovered in the F555W band.  Details given in text.
\label{fig:completeness}}
\end{figure}

We assess the completeness of our cluster catalogs by creating artificial clusters and adding them to the F555W band images. We use the Baolab task MKCMPPSF to create 4000 artificial, extended sources with an assumed King30 profile \citep{Larsen}. These artificial clusters are added randomly to the images using the MKSYNTH task, and have a range of F555W band magnitudes between 18 and 25~mag. For the artificial clusters, we adopt the approximate median size of 0.5~pixels for the detected cluster population, which we found by running Baolab on our cluser sample.

After the artificial clusters are created and added to the images, we rerun our detection and selection algorithms.  We then apply the same CI cuts that were used on the actual data. The fraction of artificial clusters of different magnitudes that are recovered by this process are plotted in Figure~\ref{fig:completeness}. The recovery fraction is quite high, above 90\%, down to a magnitude of $\approx23$.  The completeness fraction is $\approx50$\% near $m_{F555W}\sim24$ and drops rapidly below this value. 

\section{Cluster Properties}
\label{sec:analysis}

\subsection{Luminosities and Colors}
\begin{figure}[ht!]
\includegraphics[width=\linewidth]{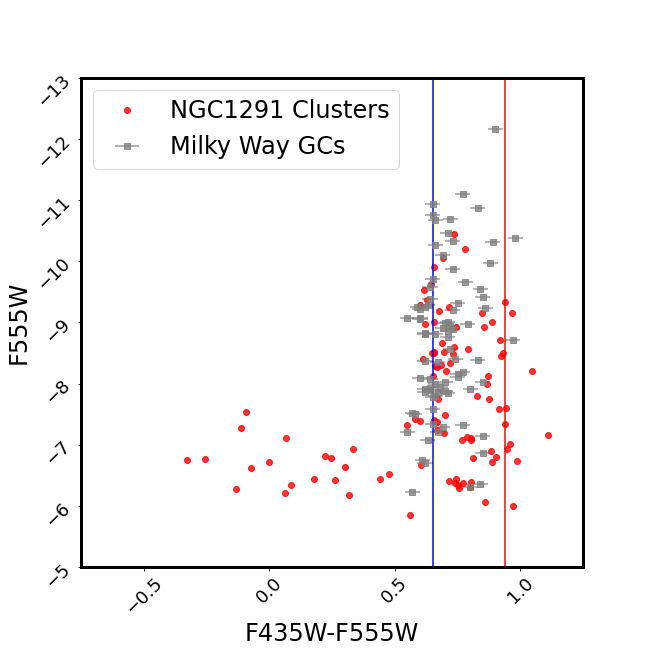}
\caption{F435W-F555W vs F555W color-magnitude diagram of all candidate clusters (red circles). Milky Way globular clusters (grey squares) from the \citet{Harris} catalog are plotted for reference with the uncertainty due to the conversion between the HST VEGAMAG and Johnson-Cousins BVI photometric systems. Magnitudes reported are absolute V band magnitudes. The blue and red lines represent the typical peak B-V color for blue, metal-poor and red, metal-rich globular clusters, respectively, for the Milky way and early-type galaxies.
\label{fig:colormag}}
\end{figure}

Figure~\ref{fig:colormag} shows the ($F435W-F555W$) versus F555W band color magnitude diagrams (CMD) of our detected cluster sample as the red circles. 
One of the most striking features in this figure is the contrast between faint, bluer clusters (with F435W-F555W$\lea 0.5$) versus the much brighter, redder clusters (F435W-F555W$>0.5$).  These bluer clusters are almost certainly young ($\lea \mbox{few}\times100$~Myr) clusters in the disk of NGC~1291, with fairly low masses.  A small population of young, low mass clusters is consistent with what we would expect from a galaxy that is currently forming stars at a low rate.  The majority of the clusters are redder, with $F435W-F555W$ colors between $\sim0.5$ and 1.0, and have magnitudes that reach up to $M_{F555W}\sim-10.5$~mag; these are candidate globular clusters in NGC~1291.  The colors and magnitudes of globular clusters in the Milky Way (gray circles; from \citet{Harris96}), shown for comparison, nicely overlap the colors and magnitudes of the redder clusters in NGC~1291.

In Figure~\ref{fig:colormag} we also show the mean ($B-V$) colors of the two color peaks found for GC systems in the Milky Way based on an updated version of the \citet{Harris96} catalog and a number of elliptical and lenticular galaxies, with typical values of 0.65 (blue, metal-poor) and 0.94 (red, metal-rich; \citep{Harris96,Kissler}). The globular clusters in NGC~1291 appear to follow this bimodal trend in color. Because colors are sensitive to metallicity in single stellar populations older than a few Gyr, the color distribution suggests that the GCs in NGC~1291 have a bimodal distribution. The color distribution is explored further in Section~\ref{sec:color_dist}.

\begin{figure}[ht!]
\includegraphics[width=\linewidth]{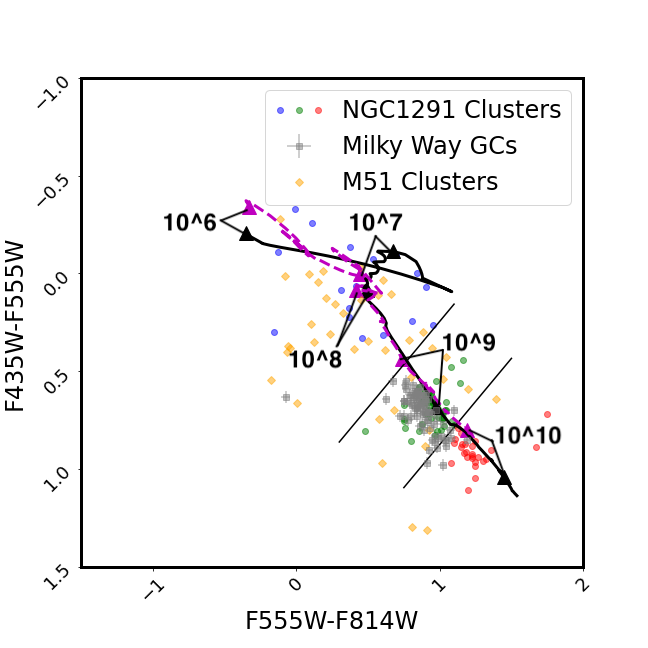}
\caption{F435W-F555W vs F555W-F814W color-color plot.  Predictions from two different cluster evolution models from \citet{BC2003} are shown, Z=0.05 (solar metallicity) as the black solid line and Z=0.004 ($\frac{1}{4} \times$ solar) as the dashed purple line.  The predicted location of clusters with ages of $10^7$, $10^8$, $10^9$, and $10^{10}$ years are indicated. Clusters in NGC~1291 follow the models closely, and are color coded by age: young clusters (blue), metal-poor globular clusters (green), metal-rich globular clusters (red).  The colors of Milky Way GCs are also shown for reference (gray),  with the uncertainty due to the conversion between the HST VEGAMAG and Johnson-Cousins BVI photometric systems. The VEGAMAG colors of bright cluster in M51 (shown in yellow) show that the blue clusters in NGC~1291 have similar colors as the young clusters formed in the disk of M51.
\label{fig:colorcolor}}
\end{figure}

In Figure~\ref{fig:colorcolor} we plot the $F435W-F555W$ vs. $F555W-F814W$ color-color diagram of clusters in NGC~1291. These are compared with the predictions from stellar evolution models from \citet{BC2003} for two different metallicities, Z=0.05 (solar), shown as a solid line, and Z=0.004 ($\frac{1}{4} \times$ solar), shown as a purple dashed line. These models start at 1~Myr (upper left) and go through ages older than 10~Gyr (lower-right), as indicated. Predictions for the evolution at the different metallicities is fairly similar, except at older ages where metal-poor clusters have bluer colors than metal-rich ones at the same age.  This is known as the age-metallicity degeneracy, and affects clusters with ages $\tau \gea 1$~Gyr.

The color-color diagram shows that there are a handful of young clusters in NGC~1291 (blue points), with ages ranging from a few to a few hundred million years. These have similar colors to bright young clusters in M51 taken from \citet{Chandar16}.
However, as already seen in the color-magnitude diagram, the majority of clusters in NGC~1291 are quite red, and overlap with the colors of Milky Way globular clusters (shown as the gray squares in Figure~\ref{fig:colorcolor}). We assume that these are ancient clusters, with ages $\gea 10$~Gyr.  The distribution of colors for these ancient GCs in NGC~1291 appears to be bimodal. To approximately define the 2 different regimes, we use two perpendicular bisectors to the stellar models, and indicate the clusters that fall into the bluer, presumably metal-poor clump (shown as green circles), and those that fall into the red, presumably metal-rich clump (red circles).



\subsection{Size Estimates}
\label{sec:sizes}

We estimate the size of each cluster using 2 different methods. The first method is the concentration index (CI), which is the difference in F555W-band magnitude measured in a 0.5 pixel and a 3 pixel aperture. The CI is a very simple measure of size that works well for most clusters.  CI is generally preferred to estimate the sizes of clusters in crowded regions because the largest aperture is only 3~pixels. Based on the CI values measured for stars and clusters in our training sets, we adopted a minimum size of $CI=2.65$ to select candidate clusters. We find that the maximum size for a cluster in our sample is $CI=3.38$, the mean value is 2.86 and the median is 2.82.

\begin{figure}[ht!]
\includegraphics[width=\linewidth]{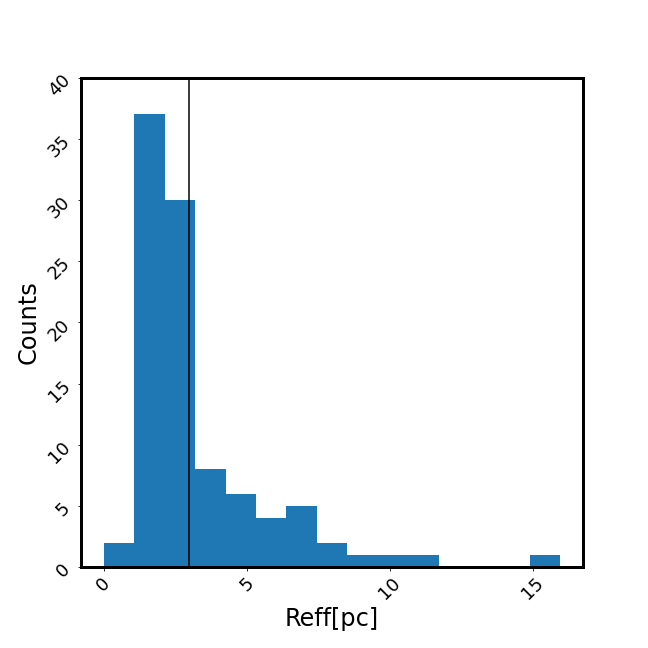} 
\caption{The distribution of measured values for $R_{eff}$ in pc for our GC sample. The median size for Milky Way GC's is overplotted as a black line.
\label{fig:Reff}}
\end{figure}

The effective radius ($R_{eff}$) of each cluster is estimated from the Ishape program in the Baolab software \cite{Larsen}. This is the radius that contains half the light. Ishape convolves analytic profiles for different $R_{eff}$, which model the profile of the surface brightness of a cluster, with the PSF to determine the best-fit size for each cluster. A subsampled PSF (as required by Baolab) was generated from $\sim30$ bright, fairly isolated point sources.  These were selected from different locations within an ACS $F555W$-band image, since this filter is the most sensitive and hence stars tend to have the highest signal-to-noise in this filter.
To run Ishape, a King 30 model is assumed for each cluster \cite{King}, although the resulting $R_{eff}$ is not sensitive to the exact profile shape that we select. Ishape returns a best-fit FWHM value  broader than the PSF in pixels, which we convert to $R_{eff}$ in parsecs using the pixel scale of $0.05\arcsec \mbox{pix}^{-1}$, assuming that $1\arcsec$ subtends 43~pc at the adopted distance of 8.9~Mpc, and including a factor of 1.48 as stated in the Ishape manual. The Ishape measurements confirm that all of our globular cluster candidates are broader than the PSF.  We find a mean FWHM ($R_{eff}$) of 1.03~pix (3.28~pc) and a median of 0.77~pix (2.46~pc), with a range between 0.31~pix (0.99~pc) and 5.0~pix (15.96~pc). These are similar to values found for the Milky Way GC system, which has a median $R_{eff}$ around 2.98~pc, and spiral galaxy M101, whose bright GCs have a median of 2.41~pc \cite{Simanton}.  The full distribution of sizes is shown in Figure~\ref{fig:Reff}.

\begin{figure}[ht!]
\includegraphics[width=\linewidth]{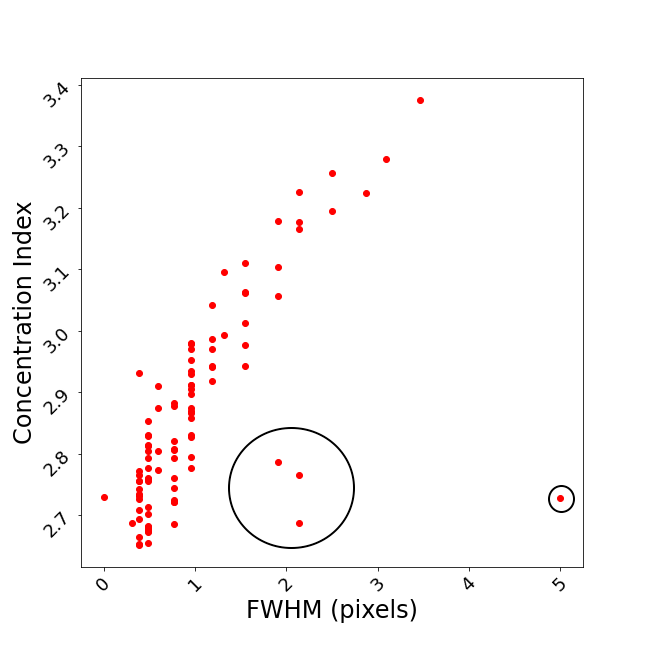} 
\caption{The measured values for CI vs FWHM from Ishape show good general agreement for the sizes of the clusters.  There are four obvious outliers (circled) that all come from crowded regions.
\label{fig:FWHMci}}
\end{figure}

We compare the relationship between CI and FWHM in Figure~\ref{fig:FWHMci} and find a strong correlation between the 2 different size measurements, although the relationship does not appear to be strictly linear (i.e. the CI values appear to "saturate"). There are four obvious outliers in this figure (circled). We find these are all in crowded regions, and therefore their FWHM values are more likely to be suspect than CI.

\section{Results} \label{sec:results}

In the rest of this paper we will focus on the ancient globular clusters (shown in green and red in Figure~\ref{fig:colorcolor}).

\subsection{Luminosity Function}
The luminosity (mass) function of the globular cluster systems in the Milky Way, M31, and elliptical galaxies have a distinctive shape, where they initially rise towards lower luminosities and masses\footnote{The luminosities of ancient globular clusters are a good proxy for their masses, although there may be some metallicity-dependence to the mass-to-light ratio.}, then flatten near $M_V\sim-7.4$~mag ($\sim10^5~M_{\odot}$) \citep{Kundu2001}; this peak magnitude is similar in the Johnson-Cousins and VEGAMAG systems. This shape is different from the one observed for young cluster populations ($<\mbox{few}\times100$~Myr), which continue to rise in power-law fashion at lower luminosities and masses \citep{Zhang}. More traditionally, when plotted as a histogram in magnitudes, GC lumniosity functions can be modeled by a Gaussian distribution that peaks near $M_V\approx-7.4$~mag. This distribution corresponds to a lognormal distibrution of their masses \citep{Jordan}. The peaked distribution for old globular clusters is believed to result from the dynamical evolution of the population, where lower mass clusters are disrupted faster than their more massive counterparts, at a given stellar density \citep{McLaughlin,Fall,Zhang}.

\begin{figure}[ht!]
\includegraphics[width=\linewidth]{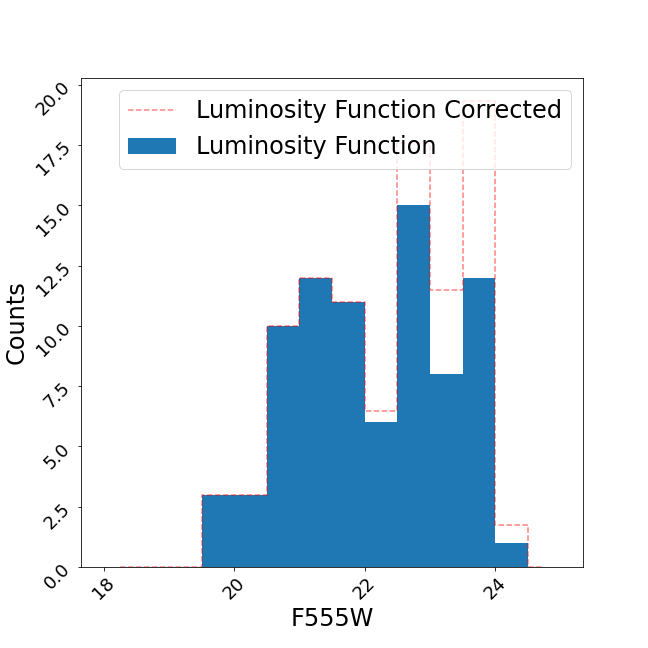}
\caption{The luminosity function for globular cluster candidates in NGC1291 is shown in blue. The red dashed lines shows the completeness corrected lumniosity function derived by boosting counts based on the recovery fraction detailed in Section~\ref{sec:completeness}.  
\label{fig:lf}}
\end{figure}

The globular cluster luminosity function in NGC~1291 is shown in Figure~\ref{fig:lf}. We see a peak in our CG luminosity function around a magnitude off 22.5, which corresponds to $M_V\approx -7.4$, similar to the peak found in the Milky Way and many other galaxies. This peak occurs above the 90\% completeness level of our sample, so is robust against incompleteness. 

\subsection{Color/Metallicity Distributions of Globular Clusters}
\label{sec:color_dist}

The metallicity distribution of GC systems sheds light on the formation history of the parent galaxy. For the GC systems in elliptical galaxies, widespread bimodality in the color (and by extension metallicity) distributions suggests that these galaxies experienced multiple epochs of cluster formation early in their histories, possibly due to mergers \citep{Kundu2001}. However, much less is known concerning the metallicity distributions of GC systems in spirals. The two best-studied spirals, the Milky Way Galaxy and M31, both have bimodal GC metallicity distributions \citep{Cote,Perrett, Huxor14}.

The colors of globular clusters are a proxy for their metallicity. 
The intrinsic metallicity distribution of GCs in the Milky Way is bimodal \citep{Cote,Perrett}. This manifests as an extended color distribution in Figure~\ref{fig:colorcolor} (gray dots). A similar extent in color is seen in the ancient cluster population of NGC~1291, as highlighted by our color coding bluer (green) vs. redder globular clusters (red).  Our clusters have mean and median F435W-F555W colors of 0.79 and 0.73, and 1.04 and 1.00 respectively, in F555W-F814W. As mentioned in Section~\ref{sec:analysis}, these colors align closely with colors of GC's in other early-type spiral galaxies and in the Milky Way.

\begin{figure}[ht!]
\includegraphics[width=\linewidth]{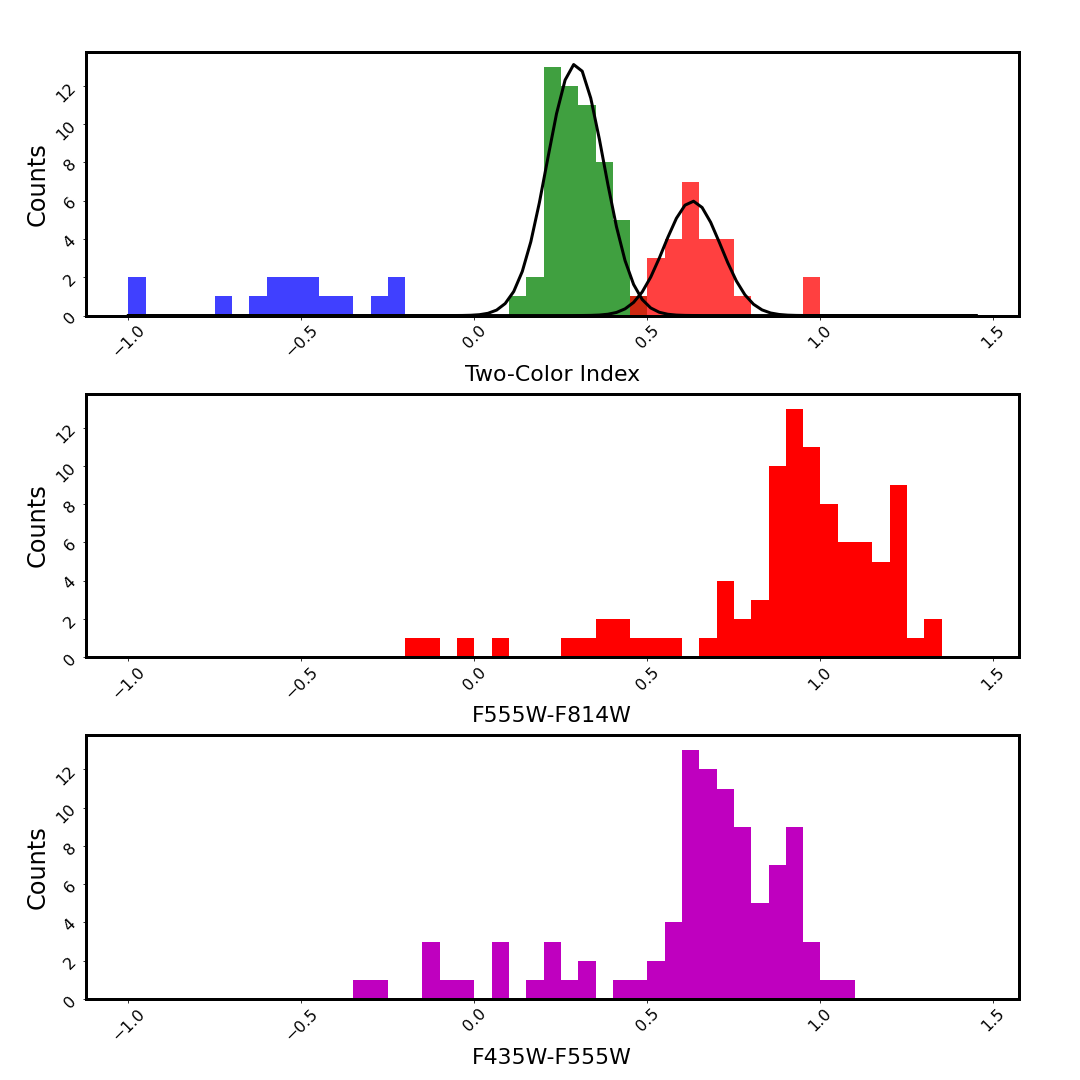}  
\caption{Histograms showing the distribution of $F555W-F814W$ (middle) and $F435W-F555W$ (bottom) colors. The top panel shows clear bimodality in the "two-color" index, where we measure the distance of each cluster along the model in $F435W-F555W$ vs $F555W-F814W$ color-color space (starting from the arbitrary zero point shown as the higher black line in Figure~\ref{fig:colorcolor}). 
Two Gaussians were fit to the two-color distribution.   
\label{fig:colorhist}}
\end{figure}

We develop a "two color index" for globular clusters by measuring their distance in color-color space (shown in Figure~\ref{fig:colorcolor}) along the $Z=0.02$ model track, starting from the perpendicular model bisector at younger ages (the results are nearly identical if we use the lower metallicity model instead). The distance of each globular cluster along the model as measured from this zero-point constitutes the "two-color index". The top panel of Figure~\ref{fig:colorhist} plots the distribution of this index for the candidate globular clusters in NGC~1291, with consistent colors between this panel and Figure~\ref{fig:colorcolor}. For comparison, we also show the distributions of single colors, with (F555W-F814W) in the middle panel and (F435W-F555W) in the bottom.  The two-color index in the top panel shows the bimodality in globular cluster colors more clearly than either of the single color indices.

The two-color index for globular clusters in NGC~1291 is well fit by a bimodal distribution, but not a unimodal one. We therefore fit two Gaussians to the distribution shown in the top panel of Figure~\ref{fig:colorhist}, using the curve\_fit function in the python package Scipy. The best fit values of each peak, height, and width $\sigma$ were returned. The blue, metal-poor (green) clusters have a mean color index located at 0.29, a peak of 13.14, and a FWHM of 0.19. The red, metal-rich (red) clusters have a mean color index of 0.63, a lower peak of 5.98, and a FWHM of 0.19. 

\subsection{Spatial Distributions}

\begin{figure}[ht!]
\includegraphics[width=\linewidth]{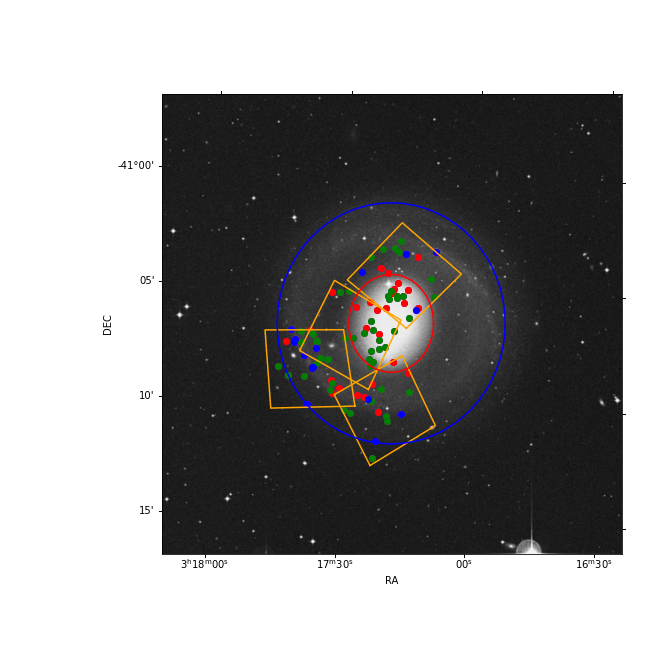} 
\caption{The locations of all selected clusters in NGC~1291 are shown on a ground-based image with the following color-coding: blue represents young, $\lea \mbox{few}\times100$~Myr clusters; green represents ancient metal-poor globular clusters, and red represents ancient metal-rich globular clusters.
The orange boxes represent our four pointings, and the inner (outer) ellipse shown in red (blue) represent our defined bulge (disk) regions, respectively. The red circle represents the bulge region and reaches out to $\sim 5.5$~kpc from the center of the galaxy, and the disk (shown by the blue circle) out to ~13.5kpc.
\label{fig:spatial}}
\end{figure}

Figure~\ref{fig:spatial} displays the location of each cluster in the galaxy on a ground based image obtained from the DSS survey \footnote{(https://archive.stsci.edu)}. The young, blue clusters appear to be located in the disk (see more below), while the red, presumably metal-rich GCs appear somewhat more centrally concentrated than the green, metal-poor GCs.  

\begin{figure}[ht!]
\includegraphics[width=\linewidth]{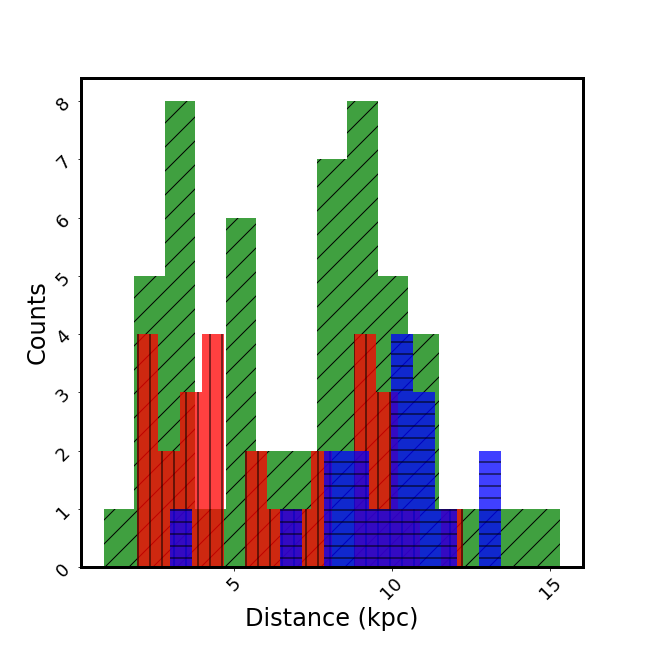} 
\caption{A histogram of the distances in kpc for young cluster (blue with horizontal hatches), metal poor GCs (green with diagonal hatches), and metal rich GCs (red with vertical hatches). Distances are calculated by projecting the face-on images onto the xy plane and finding the distance from the center of the galaxy.
\label{fig:dist_hist}}
\end{figure}

We show histogram distributions of the radial locations of the young (blue), metal-poor GC (green), and metal-rich GC (red) populations in Figure~\ref{fig:dist_hist}.  This figure shows that the young, blue clusters avoid the central region, consistent with the expectations that they reside in the disk. 
The green and red (metal-poor and metal-rich ancient GCs) are more centrally concentrated than the young clusters, consistent with expectations of spheroidal populations relative to those in galaxy disks.  While there are some gaps in the numbers because of incomplete coverage, the red clusters tend to be more evenly distributed across galactocentric distances than the green (and the coverage is identical for both populations). This is consistent with metal-rich (red) GCs being somewhat more centrally concentrated than the metal-poor (green) counterparts. This follows the expectations of a population of metal-rich GCs associated with the bulge and another of metal-poor GCs associated with the halo in NGC~1291. 


\section{The Globular Cluster Specific Frequency} \label{sec:discussion}

The specific frequency ($S_N$) allows us to compare the number of globular clusters between galaxies by normalizing the population to a galaxy of luminosity $M_V=-15$. $S_N$ is defined as:
\begin{equation} \label{Eq3}
    S_N=N_{GC}*10^{0.4*(M_V+15)}
\end{equation}
where $N_{GC}$ is the total number of globular clusters and $M_V$ is the absolute $V$-band magnitude of the galaxy \citep{Bergh}. 
An additional parameter, $T_L$ (specific mass), is sometimes used because spiral galaxies can be dominated by different age stellar populations, so their luminosities do not perfectly track their masses, which is the more fundamental parameter. $T_L$ is defined as the number of GC's per unit stellar mass ($10^9M_{\odot}$) \citep{Goudfrooij}.  Because there are fairly large uncertainties in the appropriate mass-to-light ratio to adopt for spiral galaxies, in this work we focus on specific frequency.

To calculate specific frequency and specific mass, we need the total number of clusters in NGC~1291. Along with the 81 detected globular clusters, we must account for incompleteness in our number from not including faint objects and due to incomplete coverage of the galaxy, in order to estimate the total number of globular clusters in NGC~1291.

As discussed in Section~\ref{sec:completeness}, we have a high recovery fraction, above 90\%, down to $m_{F555W}\approx23$, and in Section~\ref{sec:results}, we found that the shape of the luminosity function is similar to that of GCs in the Milky Way and other galaxies. The turnover magnitude for the Milky Way, and other early type spirals is $\approx-7.4$ \citep{Chandar}. At the distance modulus of 29.75 for NGC~1291, this corresponds to an apparent magnitude of 22.35. Since we have such high completeness brighter than the turnover magnitude, we do not apply any completeness correction for potentially missing clusters at magnitudes down to the turnover.

To correct for incomplete spatial coverage of NGC~1291, we make corrections for the bulge, disk, and outer halo regions (shown in Figure~\ref{fig:spatial}) separately. 

\begin{itemize}
\item {\em Bulge region:} we find 22 globular clusters brighter than the turnover magnitude of $M_V \approx -7.4$~mag in the bulge, and our images cover $ \approx 73\%$ of this region. We multiply by a factor of 1.27 to account for the incomplete coverage, resulting in 28 clusters brighter than the turnover. The bulge area is defined by the red ellipse in Figure~\ref{fig:spatial}.

\item {\em Disk region:} we repeat this procedure for the disk region (area between the blue and red ellipses in Figure~\ref{fig:spatial}), using a factor of 1.52 to correct for incomplete spatial coverage of the disk, and estimate a total of 64 globular clusters brighter than the turnover in this region.

\item {\em Outer halo region:} our images cover out to a galactocentric distance of $R_{gc}=13.5$~kpc.  In the Milky Way, we find only 20 globular clusters brighter than the turnover outside of this distance. We assume that the outer halo globular clusters in NGC~1291 have a radial distribution that is similar to that in the Milky Way.

\item We estimate a total of 112 globular clusters brighter than $M_V=-7.4$ by summing up the contributions from the coverage-corrected bulge, disk, and halo regions.  We then multiply this number by a factor of two to account for the clusters fainter than the turnover magnitude, resulting in an estimated total number of 224.  We assume Poisson statistics, for an {\em estimated total globular cluster population of $224\pm 15$ in NGC~1291}.

\end{itemize}

For NGC~1291 we define intrinsic properties to aid in calculations of specific frequency and mass. We adopt an absolute V band magnitude of $M_V
=-21.05$ (estimated from the total apparent $V$ band magnitude given in HyperLeda and corrected for the distance modulus of 29.75 assumed here).  
We adopt a mass of $5.8\pm{2.9}\times 10^{10} M_{\odot}$ \citep{Bittner} for NGC~1291, which includes an uncertainty of 50\% on the assumed $M/L_V$.
By using Eq~\ref{Eq3} and the stated parameters for NGC~1291, we find a specific frequency of $S_N=0.84 \pm 0.06$ and a $T_L=3.79\pm 1.90$.  If we consider the two different color populations separately, we find that the blue (red) GCs comprise 65\% (35\%) of the observed sample.  If we assume they comprise a similar fraction of the total population,  we estimate $S_{\rm N, blue} = 0.55\pm0.06$.  
If instead we estimate the total number of blue GCs by correcting for incomplete spatial coverage as described above, we find
$S_N=0.51\pm 0.06$.
Therefore, regardless of the specific method used to estimate the number of blue GCs, we find values of $S_{\rm N, blue}$ close to 0.5.
\par 
Previous studies of the GC systems in a handful of spiral galaxies have correlated their properties with galaxy type, luminosity, and mass.  One key finding was that later-type spirals (Sb and later) appear to have a fairly constant $S_N=0.5\pm 0.2$ (and $T=1.3\pm0.2$)\citep{Goudfrooij,Chandar}.
Because these galaxies are dominated by blue, metal-poor GCs, this suggests that spirals may form a fairly "universal" halo GC population. Another trend that has been suggested but not established definitively because of the poorer data quality of previous studies is that $S_N$ and $T_N$ increase from late to early-type spirals, presumably because earlier-type galaxies have formed an additional population of metal-rich GCs.  This trend appears to be unaffected by galaxy luminosity or mass.
\par 
In elliptical galaxies, the globular cluster specific frequency shows somewhat different trends, although they are also dominated by the blue GCs. Midluminosity ellipticals, which have brightnesses similar to NGC~1291, have $S_{N, blue}\approx0.5$, very similar to results for spirals of all luminosities.
However, brighter (giant) and fainter (dwarf) ellipticals have significantly higher blue GC specific frequencies, reaching values as high as 5 for some dwarf ellipticals \citep{Peng08}. The fraction of red GCs increases with luminosity, ranging from $\approx 0.1$ to $\approx 0.5$, until a luminosity of $\approx -22$. At this value, the fraction of red GC's appears to turn over \citep{Peng08}.
\par 
As a very early-type spiral (SB0/a), the GC system in NGC~1291 presents an important test of previous suggestions that spiral galaxies form universal halo GC populations.  When we only consider the blue GCs which are typically associated with halos, we found $S_{N, blue}\approx0.5$, right in line with the values found for later-type spirals that only form halo populations. Our results for NGC~1291 therefore supports the idea of a "universal" relationship between blue, metal-poor GCs and their host galaxy luminosities.
\par 
Red GCs provide clues to whether NGC~1291 better follows the trends found for spiral or elliptical galaxies.  
There are only a few spiral galaxies with estimate fractions of red GCs, which we compile below along with the galaxy type and V-band magnitude.

 {\em NGC~1291:} galaxy type$=$SB0/a, $M_V=-21.05$, $\sim35$\% red GCs  (this work)
 
{\em M81:} galaxy type$=$Sab, $M_V=-21.63$ \citep{Chandar}, $\approx40$\% red GCs \citep{Chandar} 

 {\em M31:} galaxy type$=$Sb, $M_V=-21.8$ \citep{Chandar}, $\sim34$\% red GCs  \citep{Barmby00} 
 
 {\em Milky Way:} galaxy type$=$Sbc, $M_V\approx-21.3$ \citep{Chandar}, $\sim30$\% red GCs \citep{Harris,Cote,Barmby00,Rhode}.

Spirals that are later-type than the Milky Way (Sc and later) appear to have mostly blue, metal-poor GCs (e.g., \citet{Chandar}), and hence a significantly lower fraction of red GCs than the galaxies listed above. The available information for NGC~1291, M81, M31, and the Milky Way do not show any obvious trend with galaxy luminosity, but when later-type spirals are included there is a general trend that the fraction of red GCs increases towards earlier spiral-type/bulge contribution. These trends (or lack thereof) suggests that NGC~1291 more closely follows spiral GC systems than those formed in ellipticals.
 
\par 

\section{Summary and Conclusion} \label{sec:conculsion}


\begin{itemize}
    \item We have produced a new catalog of compact star clusters 
    in the nearby ($D=8.9$~Mpc), early-type (SB0/a) spiral galaxy NGC~1291 based on $BVI$ images taken with the Hubble Space Telescope. 81 of the clusters have colors expected of ancient globular clusters, while 17 have colors expected of young ($\sim$few$\times$100~Myr) clusters. This is the first published cluster catalog in NGC~1291.
    
    
\item The ancient globular clusters have a bimodal color distribution, with $\approx65$\% ($\approx35$\%) being blue (red). The red, presumably metal-rich clusters are more centrally concentrated than the blue, presumably metal-poor, ones.

\item The correlation between color/metallicity and spatial distribution is consistent with the expectations that the red GCs are associated with the bulge, while the blue GCs are associated with the halo. The small number of young clusters are clearly located in the disk of NGC~1291.

\item The total globular cluster specific frequency is higher than that found in typical, later-type spirals. The value when only the blue globular clusters are considered, $S_{N, \rm blue}=0.50\pm0.06$, is quite similar to that for later-type spirals.  This is consistent with the expectations of a universal halo population that scales with host galaxy luminosity.  
    
\item The fraction of red GCs for NGC~1291, when compared with published results for other spiral galaxies, shows no overall trend with galaxy luminosity but is consistent with an increasing fraction from late- to early-type spirals.  These trends are dissimilar to those found for globular cluster systems in elliptical galaxies.    

\end{itemize}

\acknowledgements{We thank the anonymous referee for their comments that improved our paper.

\software{astropy (The Astropy Collaboration 2013, 2018), Baolab(Larsen 1999) }
}

\bibliography{references}{}
\bibliographystyle{aasjournal}

\section*{Tables}
\begin{deluxetable}{cccccc}[ht!]
\caption{$HST$ Pointings of NGC~1291\label{tab:pointings}}
\tablewidth{0pt}
\tablehead{
\colhead{Pointing} & \colhead{HLA Dataset} & \colhead{RA} & \colhead{DEC} & \colhead{Start Time} & \colhead{Filters} \\
\colhead{} & \colhead{} & \colhead{J2000} & \colhead{J2000} & \colhead{} & \colhead{} 
}
\startdata
PT1 & hst\_10402\_01\_acs\_wfc\_ & 3:17:16.46 & -41:04:27.8 & 2005-04-07 15:17:33 & F435W,F555W,F814W \\
PT2 & hst\_10402\_04\_acs\_wfc\_ & 3:17:28.12 & -41:07:04.1 & 2005-08-04 17:10:13 & F435W,F555W,F814W \\
PT3 & hst\_13364\_87\_acs\_wfc\_ & 3:17:37.87 & -41:08:34.0 & 2013-09-05 19:33:44 & F435W,F606W,F814W \\
PT4 & hst\_10402\_02\_acs\_wfc\_ & 3:17:19.50 & -41:10:17.7  & 2005-06-12 13:01:13 & F435W,F555W,F814W \\
\enddata
\tablecomments{Start times are different for each filter. The given Hubble data set name is not complete. The actual listing on HLA is column 2 in the table plus the corresponding filter.}
\label{tab:img}
\end{deluxetable}

\begin{center}
\begin{deluxetable}{ccccccccccc}[ht!]
\caption{Cluster Catalog\label{tab:catalog}}
\tablewidth{0pt}
\tablehead{
\colhead{ID} & \colhead{RA} & \colhead{DEC} & \colhead{F435W} & \colhead{F435W Error} & \colhead{F555W} & \colhead{F555W Error} & \colhead{F814W} & \colhead{F814W Error} & \colhead{CI} & \colhead{$R_{eff}$} \\
\colhead{} & \colhead{(J2000)} & \colhead{(J2000)} & \colhead{(mag)} & \colhead{(mag)} & \colhead{(mag)} & \colhead{(mag)} & \colhead{(mag)} & \colhead{(mag)} & \colhead{(mag)} & \colhead{(pc)} 
}
\startdata
0 & 49.312229 & -41.103973 & 21.463 & 0.009 & 20.749 & 0.008 & 19.724 & 0.008 & 2.677 & 1.533 \\
1 & 49.306077 & -41.097649 & 24.145 & 0.101 & 23.828 & 0.131 & 23.220 & 0.182 & 2.971 & 3.065 \\
2 & 49.335306 & -41.097234 & 22.873 & 0.033 & 22.008 & 0.026 & 20.905 & 0.022 & 3.041 & 3.768 \\
3 & 49.303996 & -41.096582 & 23.949 & 0.084 & 22.989 & 0.060 & 21.684 & 0.044 & 3.226 & 6.833 \\
4 & 49.318185 & -41.093568 & 23.657 & 0.067 & 22.869 & 0.056 & 21.742 & 0.048 & 2.735 & 1.245 \\
\enddata
\tablecomments{Sample data of the first 5 clusters in the catalog.}
\label{tab:cat}
\end{deluxetable}
\end{center}

\end{document}